\documentclass[conference]{IEEEtran}
\IEEEoverridecommandlockouts
\usepackage{cite}
\usepackage{balance}
\usepackage{amsmath,amssymb,amsfonts}
\usepackage{algorithm2e}
\usepackage{graphicx}
\usepackage{textcomp}
\usepackage{xcolor}
\def\BibTeX{{\rm B\kern-.05em{\sc i\kern-.025em b}\kern-.08em
    T\kern-.1667em\lower.7ex\hbox{E}\kern-.125emX}}

\usepackage{cleveref}

\usepackage[bottom]{footmisc}

\usepackage{url}
\usepackage{comment}
\usepackage{makecell}
\usepackage{float} 
\Crefname{figure}{Fig.}{Figs.}
\Crefname{section}{Sec.}{Secs.}

\begin{document}

\title{High-Resolution Channel Sounding and Parameter Estimation in Multi-Site Cellular Networks}
\author{
\IEEEauthorblockN{Junshi Chen\IEEEauthorrefmark{2}\IEEEauthorrefmark{1}, Russ Whiton\IEEEauthorrefmark{2}\IEEEauthorrefmark{3}, Xuhong Li\IEEEauthorrefmark{2}, Fredrik Tufvesson\IEEEauthorrefmark{2}}
\IEEEauthorblockA{\IEEEauthorrefmark{2}Dept. of Electrical and Information Technology, Lund University, Lund, Sweden\\
\IEEEauthorblockA{\IEEEauthorrefmark{1}Terranet AB, Lund, Sweden}
\IEEEauthorblockA{\IEEEauthorrefmark{3}Volvo Car Corporation, SE-405 31 Gothenburg, Sweden}
Email: \{junshi.chen, russell.whiton, xuhong.li, fredrik.tufvesson\}@eit.lth.se}
}
\cleardoublepage
\maketitle
\begin{abstract}
Understanding of electromagnetic propagation properties in real environments is necessary for efficient design and deployment of cellular systems. In this paper, we show a method to estimate high-resolution channel parameters with a massive antenna array in real network deployments. An antenna array mounted on a vehicle is used to receive downlink long-term evolution (LTE) reference signals from neighboring base stations (BS) with mutual interference. Delay and angular information of multipath components is estimated with a novel inter-cell interference cancellation algorithm and an extension of the RIMAX algorithm. The estimated high-resolution channel parameters are consistent with the movement pattern of the vehicle and the geometry of the environment and allow for refined channel modeling and precise cellular positioning.
\end{abstract}
\begin{IEEEkeywords}
 Channel sounding, interference cancellation, LTE, massive antenna array, RIMAX, multipath component, angle-of-arrival, delay, propagation.
\end{IEEEkeywords}
\section{Introduction}
Channel models that describe the properties of electromagnetic wave propagation in relevant environments are fundamental for the understanding and development of wireless communication systems for various purposes like communication, sensing, and positioning \cite{tataria2021standardization}. The models are typically generated through measurement campaigns performed with dedicated channel sounding hardware and high-resolution parameter estimation \cite{andreas2005rimax}.

Measurements utilizing commercial networks are appealing for a multitude of reasons. The geometry is inherently representative of real environments, transmit power and operating frequencies are not so limited as in unlicensed bands, and the commercial networks introduce the possibility to address practical issues like inter-cell interference in complex network deployments \cite{aalborg_lte_system}. 

Channel sounding systems can employ dedicated receive chains for each antenna, a single antenna moved for synthetic array creation, or fast switching among antenna elements \cite{molish2011book}. The latter is attractive for outdoor measurements because of the compromise between system complexity and measurement duration \cite{wang2019channel}. In this paper, a switched system with a 128-port stacked uniform circular antenna array is adopted. Unlike previous measurement systems for commercial base stations (BS) \cite{aalborg_lte_system, shamaei2021joint}, the system offers advantages in terms of hardware scalability, polarization diversity, and elevation resolution. 

In this work, a software-defined radio (SDR) based massive antenna channel sounding system and improvements in the associated high-resolution parameter estimation algorithm to perform channel sounding are described. The results show that the algorithms work well in real urban LTE networks with interference. 
Our research contributions are as follows
\begin{itemize}
\item Extensions of the space alternating generalized expectation-maximization (SAGE) with a maximum a-posteriori (MAP) interference cancellation algorithm that is suitable for practical environments with an unknown covariance matrix.
\item Extension of the high-resolution parameter estimation RIMAX algorithm \cite{andreas2005rimax} to improve the performance in dynamic scenarios by integrating the velocity vector of the receive antenna array.
\item We show a channel sounding system with a massive switched array that can extract high-resolution channel parameters in commercial LTE networks with interference. The analysis of the extracted parameters shows that the estimates are consistent with the movement pattern of the vehicle and the geometry of the environment. 
\end{itemize}

The signals from LTE networks are utilized in the paper because they are the ones with universal availability at the current stage,  however, the channel sounding method we proposed should be able to apply to future generation communication networks, especially for sub-6GHz networks. 

The structure of the paper is as follows. \Cref{sec:system_model} introduces the system model. \Cref{sec:sage_ic} describes the improved SAGE-MAP algorithm to cancel the inter-cell interference. \Cref{sec:rimax} explains the RIMAX algorithm incorporating the vehicle velocity information. \Cref{sec:measure} presents the measurement setup and analysis of the results from the measurement data and the proposed algorithm. Finally, \Cref{sec:summary} summarizes the paper. 

\textit{Notation}: Matrices and vectors are respectively denoted as uppercase and lowercase boldface letters, e.g., $\mathbf{A}$ and $\mathbf{a}$. $\mathbf{I}$ is an identity matrix. The superscripts $ (\cdot)^T$, $(\cdot)^H$, and $(\cdot)^{-1}$ denote matrix transpose, Hermitian transpose, and matrix inverse respectively. The operators $\odot$, $\diamond$, and $\otimes$ represent Hadamard, Khatri–Rao, and Kronecker products. The speed of light is $ c \simeq 3 \cdot 10^8$ m/s.  
\section{system model}\label{sec:system_model}
In long-term evolution (LTE) systems, the baseband signal is an orthogonal frequency division multiplexed (OFDM) signal described as
\begin{equation}
\begin{split}
r_{p,q}(t) &= \sum_{k=-N_{sc}/2}^{k=-1}{x}_{p,q}[k+N_{sc}/2]e^{j2\pi k \Delta ft}\\
&+\sum_{k=0}^{k=N_{sc}/2-1}{x}_{p,q}[k+N_{sc}/2]e^{j2\pi (k+1) \Delta ft}
\end{split}
\end{equation}
where ${x}_{p,q}[k],\, k\in \{ 0,N_\text{sc}-1 \}$ is the transmitted signal at the \(k\)-th subcarrier from the \( p\)-th antenna port of the $q$-th BS, $ p \in \{ 1,\ldots,4\} $ is the antenna port number of the cell-specific reference symbol (CRS), $q\in \{1,\ldots,Q\} $ is the number of BSs that are transmitting signals simultaneously, and \(N_{sc}\) is the number of subcarriers in the OFDM symbol. Further, $t$ is limited to $[-T_\text{CP},T_\text{s}]$ denoting continuous time, \(T_\text{CP}\) is the duration of the cyclic prefix (CP), and \( T_\text{s} = 1/\Delta f \) is
the duration of one OFDM symbol with \( \Delta f\) being the subcarrier spacing. The CRS signal $\tilde{x}_{p,q}$ is transmitted on specific subcarriers and symbols as a function of the cell ID, antenna port number, CP type, and bandwidth of the LTE system \cite{Dahlman_lte_book}. In this paper, the focus is on extracting channel parameters based on the CRS symbols, therefore, other symbols and channels are ignored.\footnote{Other synchronization signals are used for cell acquisition \cite{shamaei2018exploiting}.}

The antenna pattern of the BSs is unknown, so an assumption of omnidirectionality is applied to the transmitter antennas.\footnote{This represents a worst-case scenario, knowledge of transmitter antenna pattern could assist with channel estimation.} A 128-port stacked uniform circular antenna array is used for the channel sounder receiver. The antennas are switched in a fixed sequence with a switching interval of 0.5 ms, and all 128 ports are sampled for a complete snapshot every 75 ms. The receiver moves at a relatively low average speed of 1 m/s, which permits for treating channel parameters as constant during one snapshot, including multipath component delay, angle-of-arrival (AOA), and Doppler shift.

The frequency domain received signal model used here is similar to those in \cite{andreas2005rimax} and \cite{rui_hrpe_2017}, however, owing to the presence of multiple BSs with multiple CRS ports, the received signal of the channel sounder includes power contributions from all the CRS posts of all the BSs. It is given as follows 
\begin{equation}
    \mathbf{y}=\sum_{q=1}^Q\sum_{p=1}^P \left( {\mathbf{s}(\boldsymbol{\theta}_{sp,p,q})}+{\mathbf{n}_{dmc,p,q}} \right) \odot \mathbf{x}_{p,q} +\mathbf{n}_0.
\end{equation}
Here $\boldsymbol{s}(\boldsymbol{\theta}_{sp,p,q})$ is the specular path (SP) response, $\mathbf{n}_{dmc,p,q}$ is the diffuse multipath component (DMC) response, $\mathbf{x}_{p,q}$ is the CRS data vector, and $\mathbf{n}_0$ is the measurement noise. The parameters of the SPs are given by $\boldsymbol{\theta}_{sp,p,q}$, including both the path weight \(\boldsymbol{\gamma}\) and the structural parameter \(\boldsymbol{\mu}\), which consists of the delays \(\boldsymbol{\tau}\), azimuth AoA $\boldsymbol{\varphi}$, elevation AOA $\boldsymbol{\theta}$, and Doppler shifts \(\boldsymbol{\nu}\). $\mathbf{x}_{p,q}$ can be represented as
\begin{align}
    \mathbf{x}_{p,q} =  \left[ \tilde{\mathbf{x}}_{p,q}^T \left(1 \;\mathrm{mod}\; 20 \right),\dots, \tilde{\mathbf{x}}_{p,q}^T\left( M \;\mathrm{mod}\; 20 \right) \right]^T
\end{align}
where $M$ is the number of the receiver's antenna, and $\tilde{\mathbf{x}}_{p,q}\left(m \;\mathrm{mod}\; 20 \right)$ are all the CRS subcarriers of the $\left( m \;\mathrm{mod}\; 20 \right)$-th slot. 

Since the signals from different CRS ports are transmitted on different subcarriers, they can be separated and processed independently in the frequency domain, therefore, the receiver model of the $p$-th CRS port can be extracted as
\begin{equation}
    \mathbf{y}_p=\sum_{q=1}^Q \left( {\mathbf{s}(\boldsymbol{\theta}_{sp,p,q})}+{\mathbf{n}_{dmc,p,q}} \right)\odot \mathbf{x}_{p,q} +\mathbf{n}_0.
\end{equation}
The SP response $\mathbf{s}(\boldsymbol{\theta}_{sp,p,q})$ is given as
\begin{equation}
\begin{split}
    &\mathbf{s}(\boldsymbol{\theta}_{sp,p,q}) =  \mathbf{B} \cdot \boldsymbol{\gamma} \\
    &=\mathbf{B}_{\text{RH},p,q} \diamond \mathbf{B}_{f,p,q} \cdot \boldsymbol{\gamma}_{\text{H},p,q} 
    +\mathbf{B}_{\text{RV},p,q} \diamond \mathbf{B}_{f,p,q} \cdot \boldsymbol{\gamma}_{\text{V},p,q}\\ 
\end{split}
\end{equation}
where the basis matrices are defined as follows
\begin{align}
    \mathbf{B}_{f,p,q} &= \mathbf{G}_{f,p,q} \cdot \mathbf{A}(-\boldsymbol{\tau}_{p,q}) \\
    \mathbf{B}_{\text{RH},p,q} &= \left[ \mathbf{G}_{\text{RH}} \cdot \left( \mathbf{A}(\boldsymbol{\varphi}_{p,q}) \diamond \mathbf{A}(\boldsymbol{\theta}_{p,q}) \right) \right] \odot \mathbf{A}_{t}(\boldsymbol{\nu}_{p,q}) \label{eq:brh}\\ 
    \mathbf{B}_{\text{RV},p,q} &= \left[ \mathbf{G}_{\text{RV}} \cdot \left( \mathbf{A}(\boldsymbol{\varphi}_{p,q}) \diamond \mathbf{A}(\boldsymbol{\theta}_{p,q}) \right) \right] \odot \mathbf{A}_{t}(\boldsymbol{\nu}_{p,q}). \label{eq:brv}
\end{align}
The system frequency response $\mathbf{G}_f$ and the effective aperture distribution functions  $\mathbf{G}_\text{RH}$ and $\mathbf{G}_\text{RV}$ are available from the system and array calibration \cite{andreas2005rimax}. The phase shift matrix $\mathbf{A}(\boldsymbol{\mu}) \in \mathbb{C}^{N\times L}$ is given by
\begin{equation}
    \mathbf{A}(\boldsymbol{\mu}) = \begin{bmatrix}
  e^{-j\lfloor{\frac{N}{2}}\rfloor\mu_{1}}  & \ldots & e^{-j\lfloor{\frac{N}{2}}\rfloor\mu_{L}} \\
  \vdots &\ddots & \vdots \\
  e^{j(\lceil{\frac{N}{2}}\rceil-1)\mu_{1}} & \ldots & e^{j(\lceil{\frac{N}{2}}\rceil-1)\mu_{L}}
    \end{bmatrix}.
\end{equation}
Here $\boldsymbol{\mu}$ is a structural parameter vector that represents $\boldsymbol{\tau}_{p,q}$, $\boldsymbol{\varphi}_{p,q}$, or $\boldsymbol{\theta}_{p,q}$ which are normalized to $[ 0, 2\pi )$, $[ -\pi, \pi )$ and $[ 0, \pi]$ respectively as in \cite{andreas2005rimax}. Here $N$ is the dimension of delay, azimuth or elevation, and $L$ is the number of SPs. 

Even though the BSs are static, the movement of the vehicle and scatterers still induces Doppler shift that leads to phase rotation among the sequentially-switched receiving antennas \cite{rui_hrpe_2017}. The phase rotation is explicitly expressed as 
\begin{equation}
    \mathbf{A}_{t}(\boldsymbol{\nu}_{p,q}) = \begin{bmatrix}
  e^{j\frac{t_1}{T_0}\nu_{p,q,1}}  & \ldots & e^{j\frac{t_1}{T_0}\nu_{p,q,L}} \\
  \vdots & \ddots & \vdots \\
  e^{j\frac{t_M}{T_0}\nu_{p,q,1}} & \ldots & e^{j\frac{t_M}{T_0}\nu_{p,q,L}}
\end{bmatrix}.
\end{equation}
Here $t_m$ is the starting time of the $m$-th receive antenna. 
The velocity information from a ground truth system in the vehicle is used to facilitate the processing of the Doppler phase rotation. With the velocity information $v$, the matrix $\mathbf{A}_{t}(\boldsymbol{\nu}_{p,q})$ can be rewritten as 
\begin{equation}
\begin{split}
 \mathbf{A}_{t}(\boldsymbol{\nu}_{p,q})  = \mathbf{A}_{t}(v,\boldsymbol{\varphi}_{p,q},\boldsymbol{\theta}_{p,q})\in \mathbb{C}^{M\times L}.
\end{split}
\end{equation}
The element with the index $(m,l)$ is defined as $ e^{j\frac{ t_m }{ T_0}  f\left(v,\angle_{p,q,l}\right)} $, and $f\left(v,\angle_{p,q,l}\right)$ is a function of the antenna array velocity, the azimuth AOA and the elevation AOA of the $l$-th SP defined as 
\begin{equation}
f\left(v,\angle_{p,q,l}\right) = \frac{2\pi f_c  v T_0}{c} \cos\left( \varphi_{p,q,l}\right) \cos\left( \theta_{p,q,l}\right)
\end{equation}
where $f_c$ is the carrier frequency.

The SP weight vectors $\boldsymbol{\gamma}_{\text{H},p,q}$ and $\boldsymbol{\gamma}_{\text{V},p,q} $ are defined as follows
\begin{align}
\boldsymbol{\gamma}_{\text{H},p,q}&=b_{\text{TH},p,q} \cdot \bar{\boldsymbol{\gamma}}_{\text{HH},p,q} + b_{\text{TV},p,q} \cdot \bar{\boldsymbol{\gamma}}_{\text{VH},p,q}\\
\boldsymbol{\gamma}_{\text{V},p,q}&=b_{\text{TH},p,q} \cdot \bar{\boldsymbol{\gamma}}_{\text{HV},p,q} + b_{\text{TV},p,q} \cdot \bar{\boldsymbol{\gamma}}_{\text{VV},p,q}. 
\end{align}
Here $b_{\text{TH},p,q}$ and $b_{\text{TV},p,q}$ are the horizontal and vertical antenna response of the BS, and $\bar{\boldsymbol{\gamma}}_{\text{HH},p,q}$, $\bar{\boldsymbol{\gamma}}_{\text{VH},p,q}$, $\bar{\boldsymbol{\gamma}}_{\text{HV},p,q}$, and $\bar{\boldsymbol{\gamma}}_{\text{VV},p,q}$ are different polarization combinations of the transmitter and the receiver, e.g., $\text{HH}$ is horizontal-to-horizontal. 

The covariance matrix of the total stochastic noise can be written as
\begin{equation}
    \mathbf{R} = \mathbf{R}_{dmc}+\sigma^2_n\mathbf{I} = \mathbf{I}_t \otimes \mathbf{I}_R \otimes \mathbf{R}_f+\sigma^2_n\mathbf{I}. 
\end{equation}
Here $\mathbf{R}_{dmc}$ is the diffuse scattering model that is assumed to be independent of the measurement noise $\sigma^2_n\mathbf{I}$. The DMC follows a zero-mean complex Gaussian process with a covariance matrix having a Kronecker structure. A simplified DMC model is adopted assuming only the correlation in frequency. i.e., $\mathbf{R}_f$, and a single exponential decay model for power delay profile (PDP), and the model can be parameterized by $\boldsymbol{\theta}_d$ \cite{Richter2006_dmc}.   
\section{interference cancellation among base stations}\label{sec:sage_ic}
Inter-cell interference makes direct channel parameter estimation difficult, therefore interference cancellation is necessary to separate the signals from different BSs. In this paper, the SAGE-MAP algorithm in \cite{lte_ic} is improved and extended to the massive antenna array system. It decomposes the multi-cell channel estimation problem into single-cell channel estimation problems and accurately estimates the channel response of the serving cell and the neighboring cell with MAP/MMSE criteria.

Each antenna has a unique radiation pattern in the coordinate system of the array and will receive signals with different PDPs, so the interference cancellation is processed for each antenna separately. CRS subcarriers are used, so no interpolation of channel frequency response is applied over the time-frequency grid.

In a practical environment, the channel's Doppler power spectrum is not as ideal as the Jakes' or Gaussian model, and the delay power spectrum is not ideal as a uniform or exponential model, furthermore, these proprieties also change with the transition of scenarios from line-of-sight (LOS) to non line-of-sight (NLOS), or vice versa. For these reasons, the estimation and update of the correlation matrix of signal from each BS are based on its own estimate of the channel response. The correlation matrix is stable for a short time duration (e.g., 20 snapshots), so a simplified algorithm is applied to reduce the complexity of correlation matrix inversion considering the large amount of data measured from the massive antenna array. In the proposed algorithm, the correlation matrix is decomposed with the singular value decomposition (SVD) algorithm only once at the beginning of the iteration and the correlation matrix inversion in \cite{lte_ic} is transformed into the division of the eigenvalues and the eigenvalues plus noise.

In a real network, the time of arrival (TOA) differences between two BSs can be large. If the traditional SNR estimation algorithm is applied \cite{boumard03_snr_est}, and the TOA of the serving BS is applied to the neighboring BS, then the SNR estimation of the neighboring BS will be inaccurate due to the phase rotation caused by the delay difference. In such cases, the phase de-rotation is applied to the neighboring BS. 

The detailed interference cancellation pseudo-algorithm is described in \cref{alg:sagemapic}. The algorithm requires iteration for all the $M$ receiver antennas and all the $P$ CRS ports.
\begin{algorithm}[t]
\caption{SAGE-MAP interference cancellation algorithm}
\label{alg:sagemapic}
\LinesNumbered
\tcc{ Initialization}
\For{$q=1$ to $Q$}{
$ \mathbf{R}_{hh,m,p,q}^{t-1} = \mathbf{U}_{m,p,q}\boldsymbol{\Sigma}_{m,p,q}\mathbf{U}_{m,p,q}^H $\;
$\hat{\mathbf{h}}_{m,p,q}^{(1)} = {\mathbf{X}}_{p,q}^{-1} \left( \mathbf{y}_{m,p} - \sum\limits_{q\prime<q}\mathbf{X}_{p,q\prime}\hat{\mathbf{h}}_{m,p,q\prime}^{1}  \right) $\;
}
\tcc{ SAGE-MAP Iteration}
\For{$g=2$ to $G$}{
\For{$q=1$ to $Q$}{
$\mathbf{z}_{m,p,q} = \mathbf{y}_{m,p}-\sum\limits_{q\prime< q}\mathbf{X}_{p,q\prime}\hat{\mathbf{h}}_{m,p,q\prime}^{g}-\sum\limits_{q\prime> q}\mathbf{X}_{p,q\prime}\hat{\mathbf{h}}_{m,p,q\prime}^{g-1}$\;
$\tilde{\mathbf{h}}_{m,p,q} = \mathbf{X}_{p,q}^{-1}\mathbf{z}_{m,p,q}$\;
$\bar{\mathbf{h}}_{m,p,q} = \tilde{\mathbf{h}}_{m,p,q}\odot \mathbf{a}\left(\tau_{m,p,q}-min(\boldsymbol{\tau}_{m,p}) \right) $\; \nllabel{alg:alg_a}
$\sigma_{m,p,q}^2 = \frac{ \sum\limits_{n=1}^{N/2}\left( \bar{h}_{m,p,q}\left((n-1)\cdot 2\right) - \bar{h}_{m,p,q}\left((n-1)\cdot 2+1\right) \right)^2}{N/2}$\;
$\hat{\mathbf{h}}_{m,p,q}^{(g)}= \mathbf{U}_{m,p,q} \left( \frac{\boldsymbol{\Sigma}_{m,p,q}}{\boldsymbol{\Sigma}_{m,p,q}+\sigma_{m,p,q}^2\mathbf{I}}\right) \mathbf{U}_{m,p,q}^H\bar{\mathbf{h}}_{m,p,q} $\;
}
}
\tcc{ Update the correlation matrix}
\For{$q=1$ to $Q$}{
$ \mathbf{R}_{hh,m,p,q}^{t} = \left(1-\alpha \right)\mathbf{R}_{hh,m,p,q}^{t-1} + \alpha \hat{\mathbf{h}}_{m,p,q}^{(g)} \hat{\mathbf{h}}_{m,p,q}^{(g)H}$\; \nllabel{alg:alpha}
}
\end{algorithm}
\setlength{\textfloatsep}{2pt}

The vector $\mathbf{y}_{m,p}$ represents all the subcarriers of the received signal from the $m$-th received antenna and the $p$-th CRS port. The vector $\mathbf{h}_{m,p,q}$ represents all the subcarriers of the channel response $\mathbf{s}(\boldsymbol{\theta}_{s,p,q} ) + \mathbf{n}_{dmc,p,q}$ from the $m$-th received antenna and the $p$-th CRS port of the $q$-th BS, and the matrix $\mathbf{X}_{p,q}$ is a diagonal matrix constituted by the elements of the vector $\tilde{\mathbf{x}}_{p,q}\left(m \;\mathrm{mod}\; 20 \right)$. The vector $\mathbf{a}$ in \cref{alg:alg_a} is a phase de-rotation matrix defined as
\begin{equation}
    \mathbf{a}\left(\tau\right) = {\left[ e^{j 2 \pi f_1 \tau}, e^{j 2 \pi f_2 \tau},\ldots, e^{j 2 \pi f_N \tau} \right]} ^T
\end{equation}
here $f_n$ is the $n$-th subcarrier's frequency and the $\alpha$ in \cref{alg:alpha} is a parameter for the Alpha filter.
\section{Channel response estimation using RIMAX }\label{sec:rimax}
After the interference from other BSs is canceled, the channel model for the $p$-th CRS port of the $q$-th BS can be represented as
\begin{equation}
      \mathbf{z}_{p,q}=  {\mathbf{s}(\boldsymbol{\theta}_{sp,p,q})}+{\mathbf{n}_{dmc,p,q}} + \mathbf{n}_0.
\end{equation}
Each CRS port's signal is processed independently, so the subscripts $p$ and $q$ are dropped in the following sections. 

The distribution of $\mathbf{z}$ is $\mathcal{CN}(\mathbf{s}(\boldsymbol{\theta}_{sp}),\mathbf{R})$
 and the conditional probability is determined by
 {\small
 \begin{equation}
     \mathcal{P}_r(\mathbf{z}\vert \boldsymbol{\theta}_{sp},\boldsymbol{\theta}_d) = \frac{1}{\pi^M \det(\mathbf{R})}e^{-\left[\mathbf{z}-\mathbf{s}(\boldsymbol{\theta}_{sp})\right]^H\mathbf{R}^{-1}\left[\mathbf{z}-\mathbf{s}(\boldsymbol{\theta}_{sp})\right]}.
 \end{equation}
 }%
A RIMAX algorithm integrating the velocity information of the ground truth system is applied to the received signal. It is a robust and efficient maximum likelihood estimation (MLE) algorithm that converges faster than the SAGE algorithm \cite{sage} by jointly optimizing the parameters of all the SPs simultaneously. It also takes into account the DMC component of the channel response and gives a more stable estimation.

Accurate initial parameter estimation can help to improve the performance of the RIMAX algorithm. A classical sequential estimation method is applied in the paper. The strongest SP is detected first and subtracted from the received signal, then the second strongest SP is detected and subtracted, and so on. 

 For a single SP, the structural parameter $\boldsymbol{\mu}$ can be found by locating the peaks of the following correlation function, 
\begin{equation}
   \mathcal{C}\left(\boldsymbol{\mu},\mathbf{z}\right) = \left( \mathbf{z}^H \mathbf{R}^{-1} \mathbf{B}\right) \left( \mathbf{B}^H\mathbf{R}^{-1}\mathbf{B}\right)^{-1} \left( \mathbf{B}^H\mathbf{R}^{-1}\mathbf{z}\right)
\end{equation}
then the best linear unbiased estimator (BLUE) of $\boldsymbol{\gamma}$ is 
\begin{equation}
    \hat{\boldsymbol{\gamma}} = \left(\mathbf{B}^{H}\mathbf{R}^{-1}\mathbf{B}\right)^{-1}\mathbf{B}^{H}\mathbf{R}^{-1}\mathbf{z}.
\end{equation}
However, there is no closed-form solution to find the peak of the correlation matrix, so in chapter 5.1.2 of \cite{andreas2005rimax}, an efficient algorithm was proposed to split the multiple-dimension correlation matrix into multiple single-dimension correlation matrices and find each dimension's peak. The paper \cite{rui_hrpe_2017} further improved the performance by searching for a small space around the parameter spaces. This paper follows these methods to find the initial parameters of each SP until the residual power threshold or SP number threshold is reached.     

After all the SPs and DMC are initialized, the RIMAX algorithm alternatively optimizes the parameters $\boldsymbol{\theta}_{sp}$ and $\boldsymbol{\theta}_d$. In the paper, The simplified DMC model in \cite{andreas2005rimax} is adopted and $\boldsymbol{\theta}_d$ is estimated as described in chapter 6. The Levenberg-Marquardt method is adopted to update $\boldsymbol{\theta}_{sp}$. The $(i+1)$-th iteration's estimation is updated as 
{\small
\begin{equation}
    \hat{\mathbf{\boldsymbol{\theta}}}_{sp}^{i+1} = \hat{\mathbf{\boldsymbol{\theta}}}_{sp}^{i} + \left[ \mathcal{J}\left(\hat{\mathbf{\boldsymbol{\theta}}}_{sp}^{i},\mathbf{R} \right)+\boldsymbol{\xi}\mathbf{I}\odot \mathcal{J}\left(\hat{\mathbf{\boldsymbol{\theta}}}_{sp}^{i},\mathbf{R} \right)\right]^{-1}\mathbf{q}\left(\mathbf{z}\vert \hat{\mathbf{\boldsymbol{\theta}}}^i_{sp},\mathbf{R} \right)
\end{equation}
}%
where $\boldsymbol{\xi}$ is the update step, $\mathbf{q}\left(\mathbf{z}\vert \mathbf{\boldsymbol{\theta}}_{sp},\mathbf{R} \right)$ is the score function, and $\mathcal{J}\left(\mathbf{\boldsymbol{\theta}}_{sp},\mathbf{R}\right)$ is the Fisher information matrix (FIM). The score function and FIM are defined as follows
\begin{align}
\begin{split}
    \mathbf{q}\left(\mathbf{z}\vert \mathbf{\boldsymbol{\theta}}_{sp}, \mathbf{R} \right) &= \frac{\partial}{\partial{\boldsymbol{\theta}_{sp}}}\ln\left( \mathcal{P}_r\left(\mathbf{z}\vert\boldsymbol{\theta}_{sp},\boldsymbol{\theta}_d\right) \right) \\ 
    &=2\mathcal{R}\{ \mathbf{D}\left(\boldsymbol{\theta}_{sp}\right)^H\mathbf{R}^{-1}\left(\mathbf{z}-\mathbf{s}\left(\boldsymbol{\theta}_{sp}\right) \right) \} 
\end{split} \\
\begin{split}
    \mathbf{J}\left(\boldsymbol{\theta}_{sp},\mathbf{R} \right) & = 2\mathcal{R}\{ \mathbf{D}\left(\boldsymbol{\theta}_{sp} \right)^H \mathbf{R}^{-1}\mathbf{D}\left(\boldsymbol{\theta}_{sp}\right)\} 
\end{split}
\end{align}
where $\mathbf{D}\left(\boldsymbol{\theta}_{sp}\right)$ is the Jacobian matrix defined as
\begin{equation}    \mathbf{D}\left(\boldsymbol{\theta}_{sp}\right) = \frac{\partial}{\partial \boldsymbol{\theta}_{sp}}\mathbf{s}\left(\boldsymbol{\theta}_{sp}\right).
\end{equation}
The Jacobian matrix has many components with respect to delay, azimuth AOA, elevation AOA, and complex amplitude, and they can be represented as
\begin{align}
 \begin{split}
     \frac{\partial}{\partial \boldsymbol{\varphi}}\mathbf{s}\left(\boldsymbol{\theta}_{sp} \right) &= \mathbf{D}_\text{RH}^{\boldsymbol{\varphi}}\diamond\mathbf{B}_f \cdot \boldsymbol{\gamma}_\text{H} + \mathbf{D}_\text{RV}^{\boldsymbol{\varphi}}\diamond\mathbf{B}_f \cdot \boldsymbol{\gamma}_\text{V} 
\end{split}\\
\begin{split}
     \frac{\partial}{\partial \boldsymbol{\theta}}\mathbf{s}\left(\boldsymbol{\theta}_{sp} \right) &= \mathbf{D}_\text{RH}^{\boldsymbol{\theta}}\diamond\mathbf{B}_f \cdot \boldsymbol{\gamma}_\text{H} + \mathbf{D}_\text{RV}^{\boldsymbol{\theta}}\diamond\mathbf{B}_f \cdot \boldsymbol{\gamma}_\text{V} 
\end{split}\\
\begin{split}
     \frac{\partial}{\partial \boldsymbol{\tau}}\mathbf{s}\left(\boldsymbol{\theta}_{sp} \right) &= \mathbf{B}_\text{RH}\diamond\mathbf{D}_f \cdot \boldsymbol{\gamma}_\text{H} + \mathbf{B}_\text{RV}\diamond\mathbf{D}_f \cdot \boldsymbol{\gamma}_\text{V} 
\end{split}\\
\begin{split}
    \frac{\partial}{\partial \boldsymbol{\gamma}_\text{H}}\mathbf{s}\left(\boldsymbol{\theta}_{sp} \right) &= \mathbf{B}_\text{RH}\diamond\mathbf{B}_f
\end{split}\\
\begin{split}
    \frac{\partial}{\partial    \boldsymbol{\gamma}}_\text{V}\mathbf{s}\left(\boldsymbol{\theta}_{sp} \right) &= \mathbf{B}_\text{RV}\diamond\mathbf{B}_f
\end{split}.
\end{align}
The matrices $\mathbf{D}_\text{RH}^{\boldsymbol{\varphi}}$ and $\mathbf{D}_\text{RH}^{\boldsymbol{\theta}}$ are defined as
\begin{align}
\begin{split}
    \mathbf{D}_\text{RH}^{\boldsymbol{\varphi}} = & \left[ \mathbf{G}_\text{RH} \cdot  \left( \mathbf{D}(\boldsymbol{\varphi}) \diamond \mathbf{A}(\boldsymbol{\theta}) \right) \right] \odot \mathbf{A}_{t}(v,\boldsymbol{\varphi},\boldsymbol{\theta}) \\
    + &\left[ \mathbf{G}_\text{RH}\cdot \left( \mathbf{A}(\boldsymbol{\varphi}) \diamond \mathbf{A}(\boldsymbol{\theta}) \right) \right] \odot \mathbf{D}_{t}^{\boldsymbol{\varphi}}(v,\boldsymbol{\varphi},\boldsymbol{\theta})
\end{split} \label{eq:gda_h}\\
\begin{split}
    \mathbf{D}_\text{RH}^{\boldsymbol{\theta}} = & \left[ \mathbf{G}_\text{RH} \cdot\left( \mathbf{A}(\boldsymbol{\varphi}) \diamond \mathbf{D}(\boldsymbol{\theta}) \right) \right] \odot \mathbf{A}_{t}(v,\boldsymbol{\varphi},\boldsymbol{\theta}) \\
    + &\left[ \mathbf{G}_\text{RH} \cdot \left( \mathbf{A}(\boldsymbol{\varphi}) \diamond \mathbf{A}(\boldsymbol{\theta}) \right) \right] \odot \mathbf{D}_{t}^{\boldsymbol{\theta}}(v,\boldsymbol{\varphi},\boldsymbol{\theta}) \label{eq:gda_v}
\end{split}
\end{align}
and similarly, the matrices $\mathbf{D}_\text{RV}^{\boldsymbol{\varphi}}$ and $\mathbf{D}_\text{RV}^{\boldsymbol{\theta}}$ can be defined.

The matrices $\mathbf{D}(\boldsymbol{\varphi})$, $\mathbf{D}(\boldsymbol{\theta})$ and $\mathbf{D}_{f}$ are defined in the following matrix by replacing $\boldsymbol{\mu}$ with $\boldsymbol{\varphi}$, $\boldsymbol{\theta}$ and $\boldsymbol{\tau}$ respectively. 
\begin{equation}
\mathbf{D}(\boldsymbol{\mu}) = j\boldsymbol{\Xi} \mathbf{A}(\boldsymbol{\mu})
\end{equation}
and the matrix $\boldsymbol{\Xi}$ is a diagonal matrix with the $i$-th diagonal element being $i-1-\lfloor{N/2}\rfloor$.

$\mathbf{D}_{t}^{\boldsymbol{\varphi}}(v,\boldsymbol{\varphi},\boldsymbol{\theta})$ and $\mathbf{D}_{t}^{\boldsymbol{\theta}}(v,\boldsymbol{\varphi},\boldsymbol{\theta})$ are defined as
\begin{align}
\begin{split}
    \mathbf{D}_{t}^{\boldsymbol{\varphi}}(v,\boldsymbol{\varphi},\boldsymbol{\theta}) &= j\boldsymbol{\Psi}\odot \mathbf{A}_{t}(v,\boldsymbol{\varphi},\boldsymbol{\theta})
\end{split}\\
\begin{split}
    \mathbf{D}_{t}^{\boldsymbol{\theta}}(v,\boldsymbol{\varphi},\boldsymbol{\theta}) &= j\boldsymbol{\Phi}\odot \mathbf{A}_{t}(v,\boldsymbol{\varphi},\boldsymbol{\theta})
\end{split}
\end{align}
where $\boldsymbol{\Psi}$ and $\boldsymbol{\Phi}$ are matrices with size of $M\times L$, and the elements denoted by $(m,l)$ are $\frac{t_m}{T_0}\frac{\partial f\left(v,\angle_{l}\right)}{\partial\varphi_{l}}$ and $\frac{t_m}{T_0}\frac{\partial f\left(v,\angle_{l}\right)}{\partial\theta_{l}}$ respectively.
\section{measurement setup and data analysis}\label{sec:measure}
\begin{figure}[b]
\centerline{\includegraphics[scale=0.40]{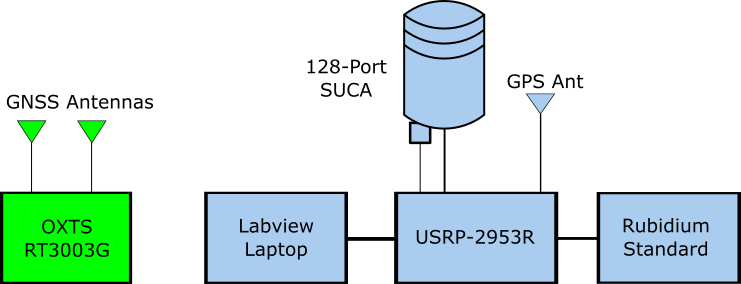}}
	\caption{Block diagram of the LTE channel sounding system, including ground truth system for precise pose estimates.}	\label{fig:Hardware_Setup}
\end{figure}
\begin{figure}	\centerline{\includegraphics[scale=0.085]{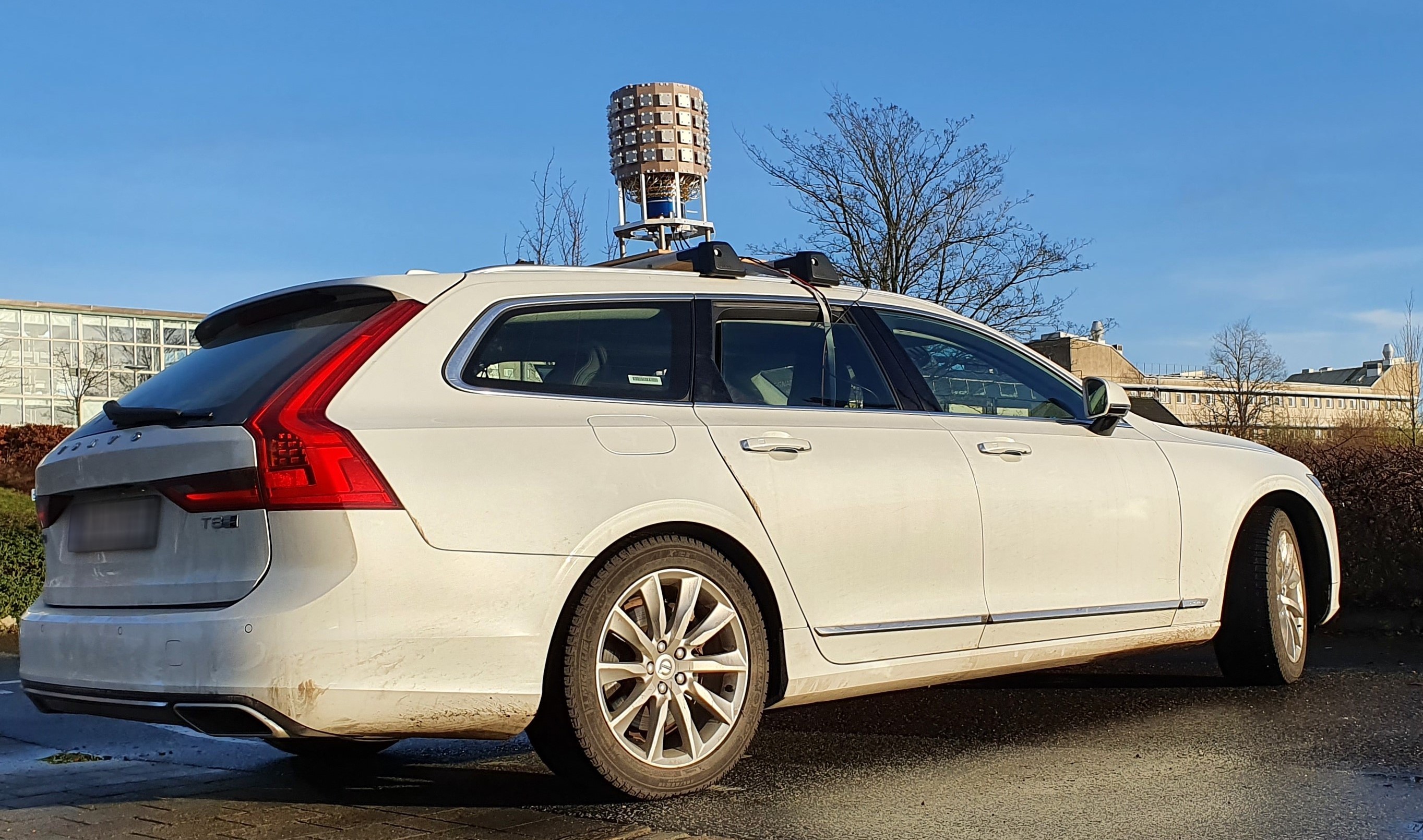}}
	\caption{The massive antenna array on top of the measurement vehicle.}
	\label{fig:volvo_car}
\end{figure}
A measurement system based on USRP-2953R and LabVIEW Communications LTE Application Framework was developed, and it was used to receive and log CRS symbols spanning 20 MHz from commercial LTE BSs operating at 2.66 GHz in the city of Lund, Sweden. A block diagram of key system components is shown in \Cref{fig:Hardware_Setup}. The USRP controlled a 128-port stacked uniform circular antenna array consisting of 4 vertically-stacked rings of 16 dual-polarized antenna elements each. The antenna array was mounted on the roof of a passenger vehicle as shown in \Cref{fig:volvo_car}. A rubidium standard disciplined by GPS was used as a frequency reference for the USRP. The OXTS RT3003G was used to provide ground truth for the position and orientation of the vehicle and the antenna array. The USRP's GPS receiver was used to synchronize the ground truth system and the USRP.
\begin{figure}
\centerline{\includegraphics[scale=0.330]{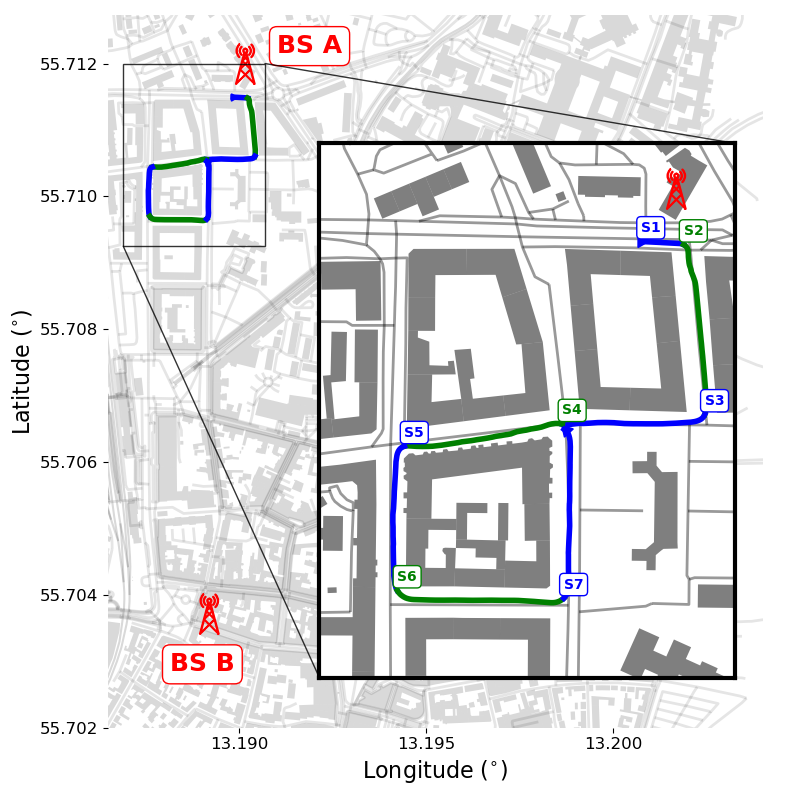}}
\caption{Measurement trajectory of the vehicle-mounted LTE channel sounding system, starting location 55.7116$^\circ$N, 13.1899$^\circ$E.}
\label{fig:mea_traj}
\vspace{-0mm}
\end{figure}

The measurement trajectory is shown in \Cref{fig:mea_traj}. The main figure gives perspective on the relative positions of the BSs and trajectory, and the inset plot gives a more granular view of the segments and BS A. During the measurement, two BSs are observed, each with three sectors. BS A includes cell IDs 375/376/377 and BS B includes cell IDs 177/178/179. These BSs are located in the north and south, and they are around 40 meters and 950 meters away from the starting point. Even though BS B is significantly further away from the receiver than BS A, the received signal strength of both BSs is comparable at some places owing to the better view and channel condition of BS B. The proposed interference cancellation algorithm can also apply to other scenarios, e.g., both BSs have similar distances to the UE.   

The vehicle's trajectory is split into 7 segments based on LOS/NLOS transitions and changes in heading. The starting time (in seconds), ending time (in seconds), the length of the segment (in meters), and the presence of the LOS signal (Yes or No) are shown in \Cref{tab:seg_time}, and the starting point of each segment is shown in \Cref{fig:mea_traj}. 
\begin{table}
  \begin{center}
    \caption{Drive segments of the measurement}
    \label{tab:seg_time}
    \begin{tabular}{|c|c|c|c|c|c|}
        \hline
        \thead{\textbf{Segment}} & \thead{\textbf{Start} \\ \textbf{(s)}} & \thead{\textbf{Stop} \\ \textbf{(s)}} & \thead{\textbf{Length} \\ (m)}  & \thead{\textbf{BS A } \\ \textbf{LOS?}} & \thead{\textbf{BS B }\\ \textbf{LOS?}} \\
        \hline
        S1 & 0 & 22 & 24 & Y & N \\
        \hline
        S2 & 22 & 128 & 98 & Y & N \\
        \hline
        S3 & 128 & 202 & 83 & N & N \\
        \hline
        S4 & 202 & 289 & 90 & N & N \\
        \hline
        S5 & 289 & 375 & 87 & N & N \\
        \hline
        S6 & 375 & 472 & 98 & N & N \\
        \hline
        S7 & 472 & 577 & 100 & N & Y\\
        \hline
    \end{tabular}
  \end{center}
\end{table}

\begin{figure}
\centerline{\includegraphics[scale=0.63]{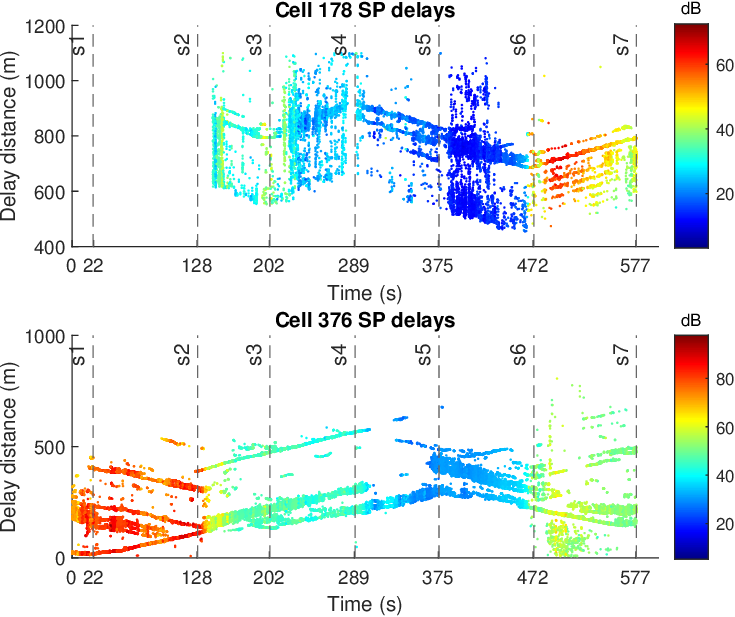}}
\caption{Specular path delay estimates for each base station after interference cancellation.}
\label{fig:mpc_delay}
\vspace{-0mm}
\end{figure}

\begin{figure}
\centerline{\includegraphics[scale=0.60]{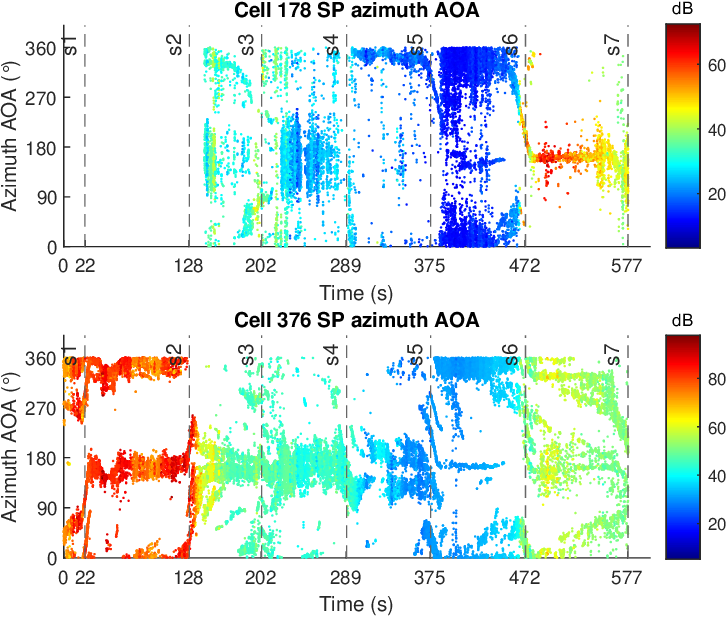}}
\caption{Specular path azimuth angle-of-arrival estimates for each base station after interference cancellation.}
\label{fig:mpc_azimuth}
\vspace{-0mm}
\end{figure}

The heat maps of the SP delay and azimuth AOA estimates from sectors 376 and 178 are shown in \Cref{fig:mpc_delay} and \Cref{fig:mpc_azimuth}. These figures are divided into 7 zones corresponding to the 7 segments, and clear delay and azimuth AOA patterns can be observed. 

During S1 and S2, the vehicle is close to BS A, and only receives signals from BS A. \Cref{fig:mpc_delay,fig:mpc_azimuth} show that the received power from BS A is high at these two segments due to the presence of the LOS path. The LOS delay increases with time because the vehicle drives away from BS A. The azimuthal angles change sharply at 22 and 128 seconds as the vehicle turns. 

The vehicle receives only NLOS signals from both BSs starting in S3 and continuing through S6, and the interference between BSs starts to become consequential. It shows that the interference cancellation algorithm can separate the signals from different BSs effectively. At S3 and S5, both BSs signals are relatively strong. At S4 and S6, the signals from BS A are strong, but the signals from BS B experience deep fades due to the urban canyons. These segments show NLOS SPs that have lifetimes of varying lengths, including two NLOS SPs with particularly long lifetimes during S3 and S4.

During S7, the vehicle receives the NLOS and LOS signals from BS A and BS B respectively, which causes the received power from BS A to be lower than that of BS B, even though BS A is closer to the vehicle. It can be observed that the interference cancellation algorithm also performs well in this scenario.  

\begin{figure}
\centerline{\includegraphics[scale=0.63]{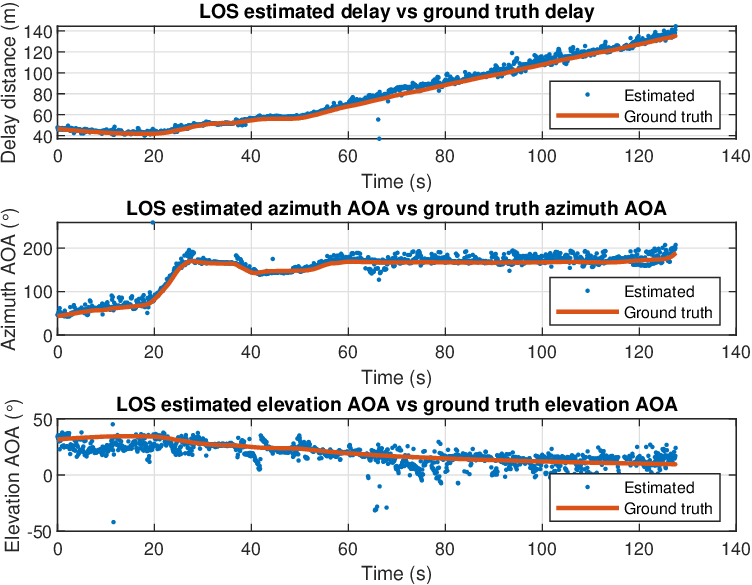}}
\caption{Comparison of parameters from the estimation and the ground truth.}
\label{fig:gt_vs_est}
\vspace{-0mm}
\end{figure}

Since the vehicle's absolute position and heading are known together with the approximate positions of the BSs, it is straightforward to calculate the ground truth delay, azimuth AOA, and elevation AOA of the LOS component. A comparison of the ground truth and the estimated parameters of the LOS during S1 and S2 from cell 376 is shown in \Cref{fig:gt_vs_est}. It can be observed that they fit well and the proposed algorithm can reach a high resolution. 
\section{conclusions}\label{sec:summary}
In this paper, a channel sounding system with a massive antenna array is developed to receive LTE signals from two commercial BSs interfering with each other. A method is proposed to first cancel the interference from the neighboring cell with the improved SAGE-MAP CRS interference cancellation algorithm, and then another method to extract specular path angular and delay parameters from these signals with a modified high-resolution RIMAX algorithm which includes the ground truth velocity to facilitate signal processing and improve estimation. Measurement results validate the algorithm, and the extracted information fits well with the movement pattern of the vehicle and the geometry of the environment. 
\section*{acknowledgement}
This work was supported by the Swedish Innovation Agency
VINNOVA through the MIMO-PAD Project (Reference number
2018-05000). Computational resources were provided by the Swedish National Infrastructure for Computing (SNIC) at HPC2N, partially funded by the Swedish Research Council through grant agreement no. 2018-05973.
The authors would like to thank Martin Nilsson for his help in setting up the measurement system.
\bibliographystyle{ieeetr}
\bibliography{MyCite}
\end{document}